\documentclass{article}
\usepackage{spconf,amsmath,graphicx}
\usepackage{cite}
\usepackage{booktabs}
\title{Speech-Mamba: Long-Context Speech Recognition with \\Selective State Spaces Models}
\name{Xiaoxue Gao and Nancy F. Chen}
\address{Institute for Infocomm Research, Agency for Science, Technology, and Research (A*STAR), Singapore}

\begin{document}
%\ninept
%
\maketitle
\begin{abstract}
Current automatic speech recognition systems struggle with modeling long speech sequences due to high quadratic complexity of Transformer-based models. Selective state space models such as Mamba has performed well on long-sequence modeling in natural language processing and computer vision tasks. However, research endeavors in speech technology tasks has been under-explored. We propose \textit{Speech-Mamba}, which incorporates selective state space modeling in Transformer neural architectures. Long sequence representations with selective state space models in \textit{Speech-Mamba} is complemented with lower-level representations from Transformer-based modeling.
\textit{Speech-mamba} achieves better capacity to model long-range dependencies, as it scales near-linearly with sequence length. 
\end{abstract}
\section{Introduction}
\label{sec:intro}

Automatic speech recognition (ASR) aims to transcribe speech into text and has garnered significant attention due to its rapid development \cite{zhang2017very,gao2024transferable,hori2017advances,luo2021simplified,dutta2022challenges}. Traditional ASR models typically employ separate implementations of acoustic, lexical, and linguistic models using a hybrid architecture \cite{hinton2012deep,sainath2013deep,xiong2018microsoft,gao2022music,povey2011kaldi}. However, recent advancements in ASR can jointly model acoustic, lexical, and linguistic components in an End-to-End (E2E) manner \cite{li2022recent,graves2014towards,gao2023self}. 
Successful E2E models include connectionist temporal classification (CTC) models \cite{graves2006connectionist}, sequence-to-sequence (S2S) models \cite{vaswani2017attention,gulati2020conformer} and the joint CTC and S2S model \cite{hori2017joint,gao2022automatic} where
the latter \cite{hori2017joint} exhibits superior performance compared to individual CTC models \cite{graves2006connectionist} and S2S models \cite{vaswani2017attention}.

Transformer architecture \cite{vaswani2017attention}, particularly in joint CTC and S2S models \cite{chang2020end,watanabe2018espnet}, has shown remarkable abilities in modeling temporal context for input sequences\cite{vaswani2017attention,gao2023polyscriber,hori2017joint}. For instance, Transformer models generally signify a substantial progression from RNN-based ASR models \cite{bahdanau2016end,graves2012sequence,chan2016listen,graves2014towards}. Specifically, Transformer-based multispeaker speech recognition models demonstrate superior performance over RNN-based models in both single-channel and multi-channel scenarios \cite{chang2020end}. The success of Transformers stems from the attention mechanism's ability to perform powerful sequence transduction by capturing the input information densely within a context window \cite{vaswani2017attention}.

\begin{figure*}[t]
\vspace{-0.4cm}
\centering
\includegraphics[width=171mm]{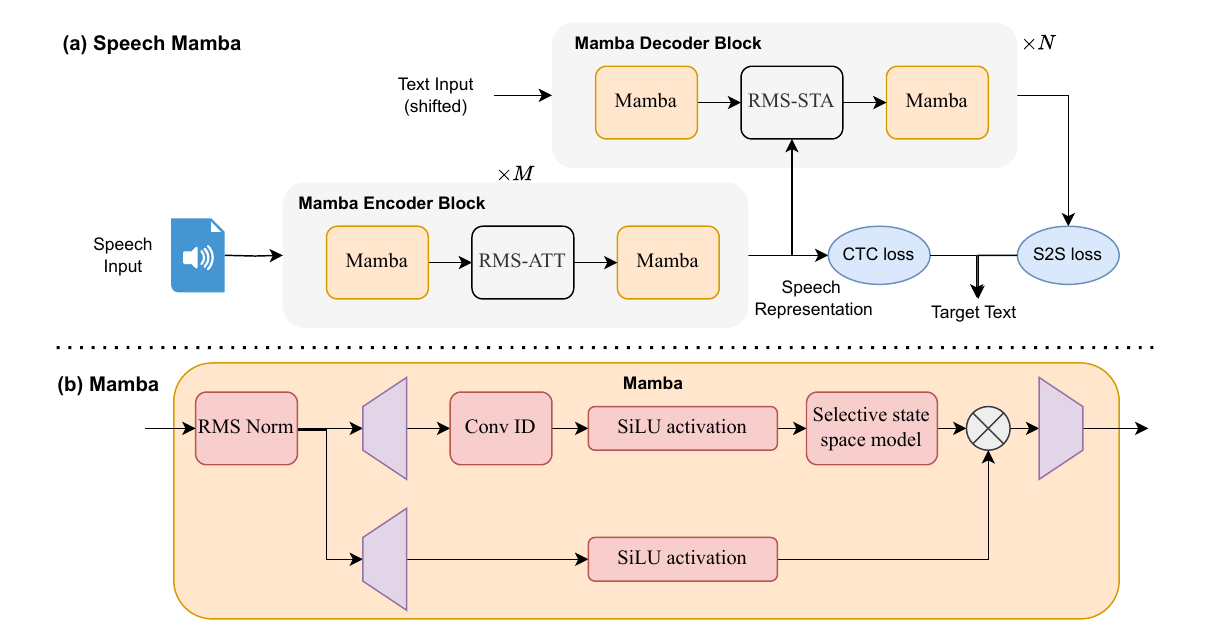}
\vspace{-0.3cm}
\caption{A framework overview of the proposed (a) \textit{Speech-mamba} approach and (b) Mamba block architecture details.}
\label{fig}
\vspace{-0.4cm}
\end{figure*}
\vspace{-0.3cm}
Recently, state space sequence models (SSMs) \cite{gu2021combining,gupta2022diagonal}, particularly structured state space sequence models (S4) \cite{gu2021efficiently}, have emerged as efficient and effective building blocks for modeling long-range dependencies in sequential data.
Mamba \cite{gu2023mamba}, as a State Space Model (SSM), has further enhanced S4 with a selective mechanism, enabling the model to choose relevant information in an input-dependent manner. Consequently, Mamba has surpassed Transformers on dense modalities and emerged as a notable approach for modeling long-range data. It excels in the field of natural language processing, characterized by its remarkable long-context modeling performance and environmentally friendly computational resources \cite{gu2023mamba}. Recognizing the advantages of Mamba, several studies in computer vision \cite{ma2024u,zhu2024vision,yang2024vivim,zheng2024u,ruan2024vm,wang2024weak,liu2024swin,wang2024semi,xing2024segmamba} and speech processing \cite{miyazaki2024exploring,zhang2024mamba} have also explored and validated its capacity in modeling long-sequence contents across various tasks.

Inspired by this success, we propose an innovative \textit{Speech-mamba}, an innovative approach for acoustic modeling in speech recognition, designed for effectively capturing long-range dependencies and achieving better transcription accuracy on long context data.
The proposed \textit{Speech-Mamba} integrates Mamba with Transformer, with Mamba specializing in capturing long-range text and speech knowledge, while Transformer focuses on modeling temporal speech and text representation.
Speech-mamba is expected to be effective in processing long sequences because of the integration of Mamba, which utilizes selective state space models with its powerful convolutional computation and near-linear computation.

Through extensive experimental validation, \textit{Speech-mamba} demonstrates its ability to mitigate the modeling challenges associated with long contexts, thereby advancing the capabilities of Transformer-style architectures in high-level speech and text representation.
By combining Mamba's strengths in capturing holistic long context with Transformer's capabilities in modeling lower-level representations, \textit{Speech-mamba} effectively addresses the challenges of long-context modeling for speech recognition and it holds promising potential as a foundational model for next-generation speech technology.

The principal contributions of this study are as follows:
\begin{itemize}
    \item We propose Speech Mamba, a novel approach that integrates selective state models with transformer neural architectures for comprehensive global and temporal context modeling in long-sequence speech recognition. 
    \item Comprehensive experiments on LibriSpeech datasets demonstrate the effectiveness of Speech Mamba in terms of recognition accuracy compared with the well-established Transformer architecture. 
    \item We propose and validate an end-to-end Mamba-based approach for long-context speech recognition.
\end{itemize}
The rest of this paper is structured as follows: Section~\ref{Speech Mamba} introduces the motivation and groundwork for this study, detailing the methodology and formulation of the proposed Speech Mamba approach. Section~\ref{Experiments} outlines the experimental setup and dataset used. Section~\ref{Results and Discussion} analyzes and discusses the experimental findings. Lastly, Section~\ref{Conclusions} provides the concluding remarks of the study.
\section{Speech Mamba}
\label{Speech Mamba}
\vspace{-0.2cm}
In this section, we present the motivation of this work, followed by the methodology and framework design of Speech Mamba.
\subsection{Motivation}
Existing speech recognition methods mostly rely on Transformer \cite{vaswani2017attention} mechanism for acoustic modeling in both CTC models and S2S models \cite{vaswani2017attention,hori2017joint,gulati2020conformer,graves2006connectionist} where Transformer excels at modeling lower-level speech and textual representation, but the self-attention mechanism poses challenges in terms of recognition accuracy capabilities when dealing with long-range text and speech dependencies. 

Mamba has demonstrated significant capability in modeling deeper speech representations, particularly after they undergo processing and compression by the encoder layers for auto-regressive speech generation \cite{gu2023mamba}. 
Inspired by this, we propose the \textit{Speech-mamba} network, which integrates Mamba into Transformer. This integration aims to leverage the strengths of both mechanisms: Transformer's robust capacity in modeling lower-level speech and textual representation and Mamba's strong capacity in modeling deeper speech and text representations.

\subsection{Mamba-integrated Speech Recognition Framework}
We propose a \textit{Speech-mamba} approach, which integrates Mamba into the acoustic model to capture deeper hidden speech representations via a Mamba encoder and hidden text knowledge alongside text embeddings through a Mamba decoder, as illustrated in Fig.\ref{fig}. 
The Speech Mamba architecture integrates a joint encoder-decoder framework with connectionist temporal classification (CTC) to convert speech input into text output
Specifically, the Mamba encoder transforms acoustic features from the speech input into intermediate hidden representations, while the Mamba decoder predicts textual sequences sequentially. This process leverages deep speech representations and previously predicted text sequences in an auto-regressive manner, illustrated in Fig.\ref{fig} (a).

\subsubsection{Methodology of \textit{Speech-mamba}}
Drawn the inspiration from Mamba that it can enhance input-dependent capacity and effectively retain crucial knowledge from long-sequence representations, we introduce a Mamba encoder to capture global long-contextual speech representation, and a Mamba decoder to learn holistic contextual cross-modal speech-text relationship for the purpose of recognition. Motivated by the effectiveness of root mean square layer normalization (RMSNorm)~\cite{zhang2019root} for Mamba modeling~\cite{lieber2024jamba}, we propose incorporating RMSNorm into the Mamba architecture, including the Mamba encoder and Mamba decoder, to stabilize the magnitude of layer activation and enhance the stability of acoustic model training, as depicted in Fig.\ref{fig} (a).

\textit{Speech-mamba} is composed of a Mamba encoder and a Mamba decoder, as depicted in Fig.\ref{fig} (a). Specifically, we incorporate $M$ Mamba encoder blocks to construct the Mamba encoder, with each block consisting of a Mamba block, followed by an RMSNorm and multi-head attention (RMS-ATT) block, and another Mamba block. Similarly, the Mamba decoder consists of $N$ Mamba decoder blocks, each containing a Mamba block, an RMSNorm and source-target multi-head attention (RMS-STA) block, and another Mamba block.

During training, the deep speech representation is obtained by jointly training the Transformer-Mamba components in the Mamba encoder with speech input. Specifically, speech acoustic features are first converted to speech representation via the initial Mamba block, capturing higher-level speech information. This representation then passes through an RMS-ATT block, followed by another Mamba block, to capture both temporal and holistic speech contexts, forming the intermediate speech representation. Subsequent Mamba encoder blocks follow the same procedure to further model these intermediate speech representations.

The Mamba decoder receives text embedding inputs derived from a text embedding layer and positional encoding operation, which converts the shifted text sequence into embedding. These text embeddings are then fed into the initial Mamba block. 
The output of the Mamba block, along with the deep speech representation, is subsequently fed into the RMS-STA block to capture the relationship between the speech representation and text embeddings.
Following this, the cross-modal representation from the RMS-STA is further modeled by the second Mamba block to learn the global text-speech knowledge. After $N$ iterations of modeling with Mamba, the deep speech and text representations are jointly compressed into smaller yet highly informative states for recognition. The residual connection and dropout are employed within both Mamba encoder and Mamba decoder. 

\subsubsection{Mamba with Selective State Models}
To capture deep long-range dependencies of text-speech features, the core module, the Mamba block, plays an important role, as illustrated in Fig.\ref{fig} (b). Mamba comprises a root mean square layer normalization (RMSNorm), a multi-layer perceptron (MLP), 1D convolution, SiLU/Swish activation function, nonlinearity operation, normalization and a selective state space model. SiLU and MLP is combined to form the gated MLP, and the input initially is normlized by RMSNorm and then traverses through the MLP and gated MLP. Subsequently, the MLP output undergoes further processing through 1D convolution, SiLU activation, and a selective state space model, resulting in the generation of hidden features. Nonlinear, MLP operations, dropout and residual connections are then applied to these hidden features and the gated MLP outputs to derive compressed yet highly representative features. 

In particular, the selective state space model (Selective SSM) plays a pivotal role in knowledge compression within Mamba, enabling the extraction of crucial contextual information necessary for modeling lengthy text and speech sequences. Acknowledging that increasing the state dimension in SSM tends to enhance the model's capacity for handling long sequences \cite{gu2023mamba}, we opt for a large state dimension. It is noteworthy that the majority of model parameters in Mamba stem from the MLP, while the contribution from the selective SSM parameters is relatively minor thanks to its powerful convolutional computation and near-linear computation. Consequently, despite the high state dimension in SSM, the model parameters remain modest, facilitating an environmentally friendly approach to speech recognition modeling.
\vspace{-0.2cm}
\subsubsection{Speech-Mamba Learning Objective}
\vspace{-0.2cm}
We employ a multi-objective learning that combines Connectionist Temporal Classification (CTC) and Sequence-to-Sequence (S2S) losses for model training. The CTC objective aids in ensuring a monotonic alignment between the input speech, encoded into a deeper acoustic representation at the output of the encoder, and the target text sequence \cite{nakatani2019improving}. Speech Mamba is trained with a combined objective function that minimizes both S2S and CTC losses simultaneously, formulated as follows:
\begin{equation}
\vspace{-0.1cm}
 \mathcal{L}_{\text{Speech-Mamba}} = \alpha \mathcal{L}^{\text{CTC}} + (1-\alpha) \mathcal{L}^{\text{S2S}} 
\label{Loss}
\vspace{-0.1cm}
\end{equation}
where $\alpha \in [0,1]$. The CTC loss is computed between encoder output after a linear transform and the target text sequence. The Mamba decoder is succeeded by linear projection and softmax layers, transforming the decoder output into a posterior probability distribution for the predicted text sequence. The S2S loss is the cross-entropy of the target text and the predicted text sequences.

\vspace{-0.2cm}
\section{Experiments}
\label{Experiments}
\subsection{Database}
\vspace{-0.2cm}
We utilize the widely-used speech recognition dataset, LibriSpeech \cite{panayotov2015librispeech}~\footnote{https://www.openslr.org/12/}, for our speech recognition experiments. We train acoustic models using standard training sets, starting with 100 hours of audio data and scaling up to 960 hours. We assess the model general recognition performance on the standard test sets, including test-other, test-clean, and dev-other subsets. Furthermore, for evaluating longer speech sequences, we construct a long context dataset where utterances are merged in sequence within per speaker to form long-context utterances exceeding 45 seconds and less than 60 seconds. The long-context datasets are curated referred to as dev-clean-L, dev-other-L, test-clean-L and test-other-L subsets, and detailed in Table~\ref{tab:datasets}. The long-context subsets will be made publicly available to support the research community in evaluating long-context speech recognition.

\begin{table}[t]
\vspace{-0.4cm}
\centering
\caption{A description of LibriSpeech dataset with utterance longer than 45 s and shorter than 60 s. We present the number of utterances, total duration and average duration for each subset.}
\label{tab:datasets}
\begin{tabular}{l|cccc}
\toprule
\textbf{Subsets} & \textbf{Total Dur (s)} & \textbf{Avg Dur (s)} & \textbf{\# Utterance}                                                  \\ \midrule
dev-clean-L       & 16960.17   &  49.30   & 344   \\ 
dev-other-L       &   16253.57 &   49.10    &331  \\ 
test-clean-L       &    16942.85&   49.54&   342    \\ 
test-other-L       &   16860.42 & 49.16 & 343        \\ 
\bottomrule
\end{tabular}
\vspace{-0.2cm}
\end{table}

\subsection{Experimental Setup}
We extract 80-dimensional Filterbank features (fbank) from audio files, and audio samples are resampled to 16k Hz.
SpeechBrain toolkit \cite{speechbrain} is used to build Transformer-based ASR baseline and the proposed Mamba-speech model.
All models are trained for 100 epochs with CTC weight $\alpha$ as 0.3 and grad accumulation factor as 4. Batch size is set to 32 with max batch length as 500 for acoustic model training. We adhere to the default procedure outlined in SpeechBrain \cite{speechbrain} of averaging the top 10 model checkpoints from the development set (dev-clean) to derive the final acoustic model.
We use pre-trained language model~\footnote{https://huggingface.co/speechbrain/asr-transformer-transformerlm-librispeech}
on LibriSpeech text for decoding for Speech Mamba and Transformer baselines. During decoding
for different ASR models, we use the same default
parameter settings (language weight, beam width and CTC decoding
weight are set to 0.6, 66 and 0.4, respectively) \cite{speechbrain}. All other parameter settings follow SpeechBrain Librispeech ASR Transformer recipe \cite{speechbrain}. 

\begin{table*}
\vspace{-0.3cm}
\centering
\caption{Comparison of the proposed \textit{Speech-mamba} and its variant Mamba-CTC models with baseline models Transformer-CTC (Trans-CTC) and Transformer for speech recognition performance (\% WER). The ASR models are trained on 100 hours of data and evaluated on standard test sets as well as long-content subsets sourced from LibriSpeech..}
\small
\begin{tabular}{lcccccccc}
\toprule
\textbf{ASR Models} & \textbf{dev-clean}& \textbf{dev-other} & \textbf{test-clean} & \textbf{test-other} & \textbf{dev-clean-L} & \textbf{dev-other-L} & \textbf{test-clean-L} & \textbf{test-other-L} \\
\midrule
Trans-CTC &9.67&	23.94&	10.42&	24.99  &30.69	&49.78&	31.69&	50.89  \\
Transformer &6.27& 15.48 &6.82 &   16.15&48.14&57.76&48.66    &58.00    \\
Mamba-CTC &6.74&	18.62	&7.38&	19.14	&7.21	&19.10&	8.07	&20.17 \\
Speech-Mamba  & \textbf{5.74} & \textbf{15.13}& \textbf{6.31}&\textbf{15.93}& 7.45&\textbf{18.47}  &\textbf{7.71} & \textbf{19.48} \\
\bottomrule
\end{tabular}
\label{main}
\end{table*}

\subsection{Model Architecture and Baselines}
Speech Mamba consists of one Mamba encoder and one Mamba decoder where Mamba encoder include seven Mamba encoder blocks ($M$ is set to 7). Mamba decoder includes three Mamba decoder blocks ($M$ is set to 3). In RMS-STA and RMS-ATT, attention dim is 512, the number of heads is 8, as in SpeechBrain Librispeech ASR Transformer recipe \cite{speechbrain}. In the Mamba block, we configure the local convolution width as 4, the model dimension as 512, the SSM state dimension as 256, the expansion factor as 2.

We use a Transformer ASR model with joint CTC and S2S losses as our baseline. This model consists of twelve encoder blocks and six decoder blocks, with an attention dimension of 512, eight heads, and a feedforward network (FFN) layer dimension of 2,048, as specified in the SpeechBrain Librispeech ASR Transformer recipe \cite{speechbrain}.
To investigate the impact of employing a multi-objective function with the Mamba block, we remove the S2S loss, creating a variant called Mamba-CTC for our ablation study. Similarly, we create a variant of the Transformer model, Trans-CTC, by removing the S2S loss, allowing us to compare it directly with Mamba-CTC.

\begin{table*}
\vspace{-0.3cm}
\centering
\caption{Ablation study of the proposed Speech-Mamba on speech recognition performance (WER \%). The ASR models are evaluated on both standard test sets and long-content subsets sourced from LibriSpeech.}
\small
\begin{tabular}{lcccccccc}
\toprule
\textbf{ASR Models} & \textbf{dev-clean}& \textbf{dev-other} & \textbf{test-clean} & \textbf{test-other} & \textbf{dev-clean-L} & \textbf{dev-other-L} & \textbf{test-clean-L} & \textbf{test-other-L} \\
\midrule
Speech-Mamba  & 5.74 & 15.13& 6.31&15.93& 7.45&18.47  &7.71 & 19.48 \\\midrule
--Mamba encoder&6.38	&15.54&	7.01	&15.78
& 45.35&	57.09&	46.57	&57.33 \\
--Mamba decoder&5.89&	15.20&	6.33	&15.90& 19.10 &33.06	&20.33	&34.06  \\
--Multi-objective&6.74&	18.62	&7.38&	19.14	&7.21	&19.10&	8.07	&20.17 \\
\bottomrule
\end{tabular}
\label{ablation}
\vspace{-0.3cm}
\end{table*}
\vspace{-0.2cm}
\section{Results and Discussion}
\label{Results and Discussion}
We investigate the general recognition performance, the impact of multi-objective learning and the effectiveness of modeling long sequences. Additionally, we conduct an in-depth ablation study to comprehensively assess the impact of using Mamba across various aspects of speech recognition. Scaling up the model, we compare its performance against state-of-the-art systems in speech recognition. Our evaluation reports transcription performance using word error rate (WER), calculated as the ratio of total insertions, substitutions, and deletions to the total number of words

\subsection{General Recognition Performance}

To evaluate the overall recognition performance, we initially assess the speech recognition capabilities of the proposed model, which is trained on 100 hours of audio and tested on standard test sets. These test sets include both short and long utterances from LibriSpeech (dev-other, test-clean, and test-other), as shown in Table~\ref{main}. Remarkably, our proposed Speech-Mamba model outperforms the Transformer baseline, and the Mamba-CTC variant outperforms Trans-CTC. This demonstrates the effectiveness of the Mamba-based model for speech recognition.

\subsection{Effect of Multi-objective Learning}
To investigate the impact of employing multi-objective learning with \textit{Speech-mamba}, we compare its performance against Mamba-CTC without multi-objective learning in Table~\ref{main}. The results demonstrate that \textit{Speech-mamba} generally achieves better performance across both short and long test sets compared to Mamba-CTC. This highlights the potential of multi-objective learning to enhance the capabilities of Mamba for speech recognition.

\begin{table}
\vspace{-0.3cm}
\centering
\caption{Comparison between the proposed \textit{Speech-mamba} model and Transformer baseline on longer-context testsets created from LibriSpeech dataset on speech recognition performance (WER \%).}
\small
\label{long-context}
\begin{tabular}{l|cc}
\toprule
\textbf{Models} & \textbf{Transformer}& \textbf{Speech-Mamba} \\ \midrule
\textbf{dev-clean-70} & 55.61& 8.23\\
% \textbf{dev-other-70} & r5& 19.87\\
% \textbf{test-clean-70} &r5 &r5\\
% \textbf{test-other-70} &r5 &rz  \\
%  \midrule
 \textbf{dev-clean-80} &61.06 &\textbf{8.69} \\
\textbf{dev-clean-90} & 64.31 &9.57 \\
% \textbf{test-clean-80} & r5/r4&9.10\\
% \textbf{test-other-80} & & 21.08 \\
\textbf{dev-clean-100} & 67.87& \textbf{10.74} \\
% \textbf{dev-other-100} & r4& 23.79\\
% \textbf{test-clean-100} & r4&11.04\\
% \textbf{test-other-100} & r4& 24.35\\
\bottomrule
\end{tabular}
\vspace{-0.4cm}
\end{table}

\begin{table*}
\vspace{-0.4cm}
\centering
\caption{Comparison between the proposed Speech-Mamba framework and other existing solutions on standard and long-context testsets using LibriSpeech dataset on speech recognition performance (WER \%).}
\small
\label{SOTA}
\begin{tabular}{l|cccc}
\toprule
\textbf{Models} & \textbf{Gemini-1.5-pro} & \textbf{Whisper-Large-V3} & \textbf{Transformer}& \textbf{Speech-Mamba} \\ \midrule
\textbf{Training Data}&-& 5 million hours& 960 hours& 960 hours\\
\textbf{Model Parameters}&- &1550 M& 71.5 M & 67.6 M\\\midrule
\textbf{dev-clean-L} &3.14 &9.33 & 28.73 &  \textbf{2.59}\\
\textbf{dev-other-L} &4.79 &7.09& 35.99 & 6.36 \\
\textbf{test-clean-L} &3.27 & 9.42& 29.24&  \textbf{2.81}\\
\textbf{test-other-L} & -&7.08 & 37.32 & \textbf{6.55}
\\\midrule
\textbf{dev-clean} & 4.65&11.32&2.09 &2.17 \\
\textbf{dev-other} & 7.19&13.22&4.89 & 5.16 \\
\textbf{test-clean} &4.92& 11.18& 2.36& 2.34\\
\textbf{test-other} & 7.41&12.86&5.31 & 5.53\\
\bottomrule
\end{tabular}
\vspace{-0.4cm}
\end{table*}
\subsection{Effectiveness of Modeling Long Sequence}
To evaluate the ability to model longer sequences, we assess the speech recognition performance of the proposed models and baselines on long-context utterances exceeding 45 seconds but less than 60 seconds (dev-other-L, test-clean-L, and test-other-L), detailed in Table~\ref{main}. 
\textit{Speech-mamba} shows significant improvements over the Transformer baseline, achieving relative improvements of over 65\% across all subsets, notably reaching relative improvements of 84\% for test-clean-L and dev-clean-L.
These improvements are more pronounced in clean test sets due to the training data's clean nature. Similarly, Mamba-CTC outperforms Trans-CTC and the Transformer baseline across all long-context test sets. This underscores the effectiveness of the \textit{Speech-mamba} framework in modeling longer sequences.

% To further investigate the impact of modeling longer speech utterances, we expand our evaluation of \textit{Speech-mamba} and the Transformer baseline using extended context test sets derived from standard LibriSpeech data. We construct subsets such as dev-clean-80 by concatenating utterances sequentially per speaker to create long-context utterances lasting between 55 and 70 seconds. Similarly, we form subsets with durations ranging from more than 65 seconds to less than 80 seconds (dev-clean-80, dev-other-80, test-clean-80, and test-other-80), and from more than 85 seconds to less than 100 seconds (dev-clean-100, dev-other-100, test-clean-100, and test-other-100). 

To further investigate the impact of modeling longer speech utterances, we expand our evaluation of \textit{Speech-mamba} and the Transformer baseline using extended context test sets derived from standard LibriSpeech data. We construct subsets such as dev-clean-80 by concatenating utterances sequentially per speaker to create long-context utterances lasting between 65 and 80 seconds. Additionally, we create subsets with durations of 55 to 70 seconds (dev-clean-70), 75 to 90 seconds (dev-clean-90), and 85 to 100 seconds (dev-clean-100).

These subsets are designed to evaluate the performance of the proposed \textit{Speech-mamba} model across varying lengths of speech utterances, as detailed in Table~\ref{long-context}. 
It is clear from Table~\ref{long-context} that \textit{Speech-mamba} consistently surpasses the Transformer baseline across utterances of different lengths, underscoring its effectiveness in modeling long sequences of data.
\vspace{-0.2cm}
\subsection{Ablation Study}
To gain deeper insights into the \textit{Speech-mamba} model's contributions, we conducted an ablation study employing three distinct approaches detailed in Table~\ref{ablation}. Initially, we substituted the Mamba encoder with a Transformer encoder identical to the one used in our Transformer baseline (second row - Mamba encoder in Table~\ref{ablation}). The results revealed a significant performance decline, particularly noticeable with lengthy sequences. This underscores the pivotal role of the Mamba encoder in effectively capturing long speech representations from extensive speech contexts. 

Next, we replaced the Mamba decoder from \textit{Speech-mamba} with a Transformer decoder, consistent with the Transformer baseline (third row - Mamba decoder in Table~\ref{ablation}). Similarly, we observed a performance decrease when replacing the Mamba decoder, though less pronounced compared to substituting the Mamba encoder with long-sequence data. This suggests that while the Mamba encoder holds greater importance in modeling extended sequences, the Mamba decoder remains crucial for capturing long-context textual data and maintaining cross-modal speech-text relationships.

We then remove the multi-objective learning objectives, leaving only the CTC loss (fourth row - Multi-objective in Table~\ref{ablation}), and observe a general performance decrease across both short-term and long-term test sets. This further confirms the effectiveness of incorporating multi-objective learning in the training of the \textit{Speech-mamba} model.

\vspace{-0.2cm}	
\subsection{Comparison with the State-of-the-Art}
We compare the proposed Speech Mamba with several state-of-the-art models, including the Transformer ASR in SpeechBrain \cite{speechbrain}, Gemini-1.5-Pro~\cite{reid2024gemini}\footnote{gemini-1.5-pro-preview-0514} and Whisper-Large-V3~\cite{radford2022whisper}\footnote{https://huggingface.co/openai/whisper-large-v3} in Table~\ref{SOTA} for both standard and long-context testsets on Librispeech. Gemini 1.5 Pro is notable as the latest large multimodal model and excels particularly in processing long-context data across text, video, and audio modalities, especially in long-context speech recognition, which aligns with the objectives of this work~\cite{reid2024gemini}. Whisper-Large-V3 is a powerful state-of-the-art model in the speech recognition domain, trained on 1 million hours of weakly labeled audio and 4 million hours of pseudolabeled audio for both speech recognition and speech translation~\cite{radford2022whisper}

To ensure a fairer comparison with the powerful Gemini and Whisper models, we scale up both the Transformer baseline and the proposed Speech Mamba model, training them on 960 hours of audio from LibriSpeech. We do not report test-other-L results for Gemini, as over 10\% of the data could not be predicted due to Gemini's safety filter.
We first examine the trainable model parameters to study the effect of model size. Our findings reveal that Speech Mamba requires fewer training parameters compared to the Transformer baseline and Whisper-Large-V3, suggesting its potential for more environmentally friendly model development.

We can observe that the Transformer performs well on normal sentences but struggles with long contexts. In contrast, Speech Mamba matches the Transformer on
short contexts and outperforms it on long-context data. This underscores the advantages of the Mamba architecture over purely Transformer-based models. 
Speech Mamba also surpasses Gemini on the test-clean-L and dev-clean-L sets and achieves competitive performance on the dev-other-L set. Note that Gemini's performance in these comparisons should appear worse than in Table~\ref{SOTA}, as we regard ground-truth text as Gemini's predictions to calculate WER for the 24 utterances that Gemini could not predict due to its safety filter. (normal: dev-clean:46; dev-other:73; test-clean:52; test-other107)
This further confirms Speech Mamba's effectiveness in long-sequence modeling. Our proposed Speech Mamba also consistently outperforms the Whisper-Large-V3 model across all test sets, despite Whisper-Large-V3 being trained on 1 million hours of audio data compared to our use of only 960 hours, and Whisper-Large-V3 having significantly more model parameters (1550M) compared to our model's 67.6M.

\vspace{-0.2cm}
\section{Conclusions}
\vspace{-0.3cm}
\label{Conclusions}
We propose a novel approach, \textit{\textit{Speech-mamba}}, which incorporates selective state models into transformer for long-context speech recognition. \textit{Speech-mamba} successfully leverages Mamba's capacity for learning representations from long sequences and Transformer's ability to model lower-level temporal knowledge.
\textit{Speech-mamba} serves as an important step into the exploration of long-context speech recognition via an E2E Mamba-integrated design.
Experiments conducted on the LibriSpeech showcase the efficacy of \textit{Speech-mamba} in recognizing long sequences of speech compared with transformer. Future endeavors will involve applying \textit{Speech-mamba} to other languages and speech processing tasks. Codes can be accessed at the link \footnote{https://github.com/xiaoxue1117/speech-mamba-public}.

\vspace{-0.2cm}

\section{ACKNOWLEDGMENTS}
This research is supported by the National Research Foundation, Singapore under its AI Singapore Programme (AISG Award No: AISG2-GC-2022-005). This project is also supported by Ministry of Digital Development and Information (MDDI) and National Research Foundation (NRF), under the Public Sector Translational R\&D Grant Funding Initiative (TRANSGrant). The aim of the funding initiative is to tap on the research community to solve public sector challenges with innovative use of digital technologies. Any opinions, findings and conclusions or recommendations expressed in this material are those of the author(s) and do not reflect the views of National Research Foundation, Singapore.

\footnotesize
\bibliographystyle{IEEEbib}
\bibliography{refs}
\end{document}